\documentclass[aps,pra,showpacs,preprint]{revtex4-1}
\usepackage{amsmath}
\usepackage{graphicx}
\usepackage{dcolumn}
\usepackage{bm}
\usepackage{textcomp}
\usepackage{natbib}
\usepackage{upgreek}
\usepackage{enumerate}

\def\ket#1{\mathinner{|{#1}\rangle}}

\begin{document}
\title{Coherent population pumping in a bright state}
	
\author{Sumanta Khan}
\author{Vineet Bharti}
\author{Vasant Natarajan}
\email{vasant@physics.iisc.ernet.in}
	
\affiliation{Department of Physics, Indian Institute of Science, Bangalore-560012, India}

\begin{abstract}
We demonstrate resonances due to coherent population pumping in a bright state (CBS), using magnetic sublevels of the closed  $ F_g = 2 \rightarrow F_e = 3 $ transition in $^{87}$Rb. The experiments are performed at room temperature vapor in two kinds of cells---one that is pure and the second that contains a buffer gas of Ne at 20 torr.  We also present the effect of pump power variation on the CBS linewidth, and explain the behavior by using a power-dependent scattering rate. The experimentally observed CBS resonances are supported by a density-matrix analysis of the system. 
\end{abstract}
	
	\maketitle	

\thispagestyle{empty}

\section*{Introduction}
Coherent population trapping (CPT) is a well studied phenomenon in many atoms. It is a phenomenon in which atoms get optically pumped into a dark non-absorbing state by two phase-coherent beams. Once they are pumped, the atoms get trapped in the dark state and cannot fluoresce because the coupling of this superposition state to the excited state cancels \cite{ARI96}. The easiest way to observe this experimentally is to use magnetic sublevels of a degenerate transition. The required phase coherence is then achieved by deriving both beams from the same laser. A narrow absorption dip then appears at line center when one of the two beams is scanned; the line center being the point at which the two-photon Raman resonance condition is satisfied. The linewidth of the dip is much smaller than the natural linewidth of the excited state, and is limited by decoherence among the ground sublevels.

A similar arrangement with two phase-coherent beams can be used to create a bright superposition state. The result is enhanced absorption at line center, exactly opposite to the dip seen in CPT. The linewidth is similar to that obtained in CPT, and is again limited by decoherence among the magnetic sublevels of the ground state. However, unlike in CPT, the population does not get trapped in this state because it can decay by coupling to the excited state. The conditions for observing this in a $F_g \rightarrow F_e $ transition are:
\begin{itemize}
	\item[(i)] It is a closed transition, so that there is no decay out of the system.
	\item[(ii)] $ F_e = F_g + 1 $, so that the correct superposition state can be formed.
	\item[(iii)] $ F_g  \neq 0 $, so that there are multiple magnetic sublevels in the ground state. 
\end{itemize}
All these conditions are met for the $ F_g = 3 \rightarrow F_e = 4 $ transition in $^{85}\rm Rb $, which was therefore used for the first observation of such increased absorption in Ref.~\cite{LBA99,LBL99}. The authors called the phenomenon electromagnetically induced absorption (EIA), in order to highlight the fact that there was increased absorption at line center. However, we feel that a more appropriate term would be CBS standing for coherent population pumping in a bright state, while the term EIA is better used for enhanced absorption of a weak probe beam in the presence of two or more strong pump beams in a multilevel system \cite{GWR04,KTW07,BMW09,CPN12,BWN16}.

In this work, we study CBS resonance satisfying the above conditions but in the other isotope of Rb, namely the $ F_g = 2 \rightarrow F_e = 3 $ transition in $^{87}$Rb. We experimentally study these resonances in two kinds of vapor cells---one that is pure and contains both isotopes in their natural abundances, and the second that contains only $^{87}$Rb and has a buffer gas of Ne at 20 torr. The presence of the buffer gas is advantageous because it increases the coherence time among the magnetic sublevels, and hence results in a smaller linewidth for the resonance. The explanation of enhanced absorption at line center for this transition is borne out by a numerical density-matrix analysis, which takes into account Doppler averaging in room temperature vapor. We also study the effect of power variation on the linewidth of the CBS resonances, and find that it follows the power-dependent scattering rate from the excited state.

\section{Experimental details}
\begin{figure}[!h]
 	\centering
 	\includegraphics[width=.8\textwidth]{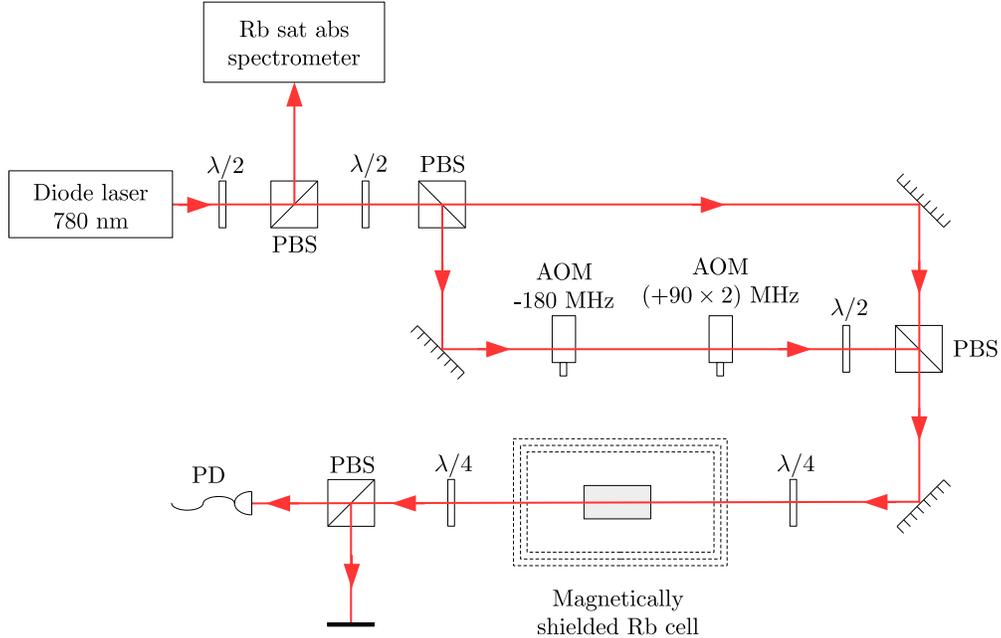}
 	\caption{(Color online) Experimental setup for CBS experiment. The required phase coherence is achieved by deriving both beams from a single laser. The probe beam is locked while the pump beam is scanned by scanning the frequency of the double-passed AOM. Figure key: $\lambda/2$ -- half wave retardation plate; $\lambda/4$ -- quarter wave retardation plate; PBS -- polarizing beam splitter cube; AOM -- acousto-optic modulator; PD -- photodiode.}
 	\label{cbsschematic}
 \end{figure}

The experimental setup is shown schematically in Fig.\ \ref{cbsschematic}. The probe and pump beams are derived from the same laser to achieve the required phase coherence. The laser consists of a grating stabilized diode laser system, as described in reference \cite{MRS15}. The linewidth of the laser after stabilization is 1 MHz. The size of the output beam is 3 mm $\times$ 4 mm. The power in the beams is controlled using $\lambda/2$ waveplates in front of the respective PBSs. 

The two beams are made to have orthogonal linear polarizations so that they can be mixed on a PBS. The experiment requires them to have circular polarizations, which is achieved by using a $ \lambda/4 $ waveplate before entering the cell. The laser is locked to the $ F_g = 2 \rightarrow F_e = 3 $ transition using a saturated absorption (SAS) signal from another vapor cell. The orthogonal circular polarizations for the two beams means that the probe beam couples $ m_{F_g} \rightarrow m_{F_g} + 1 $ transitions, while the pump beams couples $ m_{F_g} \rightarrow m_{F_g} - 1 $ transitions. As mentioned before, the probe beam frequency is fixed while that of the pump beam is scanned. This scanning is achieved by using two AOMs in its path---one with a downshift of 180 MHz, and the other compensating for this shift by a double-passed AOM with an upshift of 90 MHz. The double passing ensures that the direction of the beam does not change when the frequency is scanned. The frequency of the AOM driver is set using a commercial function generator.

Two kinds of vapor cells were used for the experiment---one pure and the second with a buffer gas of Ne (at a pressure of 20 torr). Both cells are cylindrical with dimensions of 25 mm diameter $ \times $ 50 mm length. The cell is inside a 3-layer $ \upmu $-metal magnetic shield.  The shield reduces stray external fields to less than 1 mG.

The polarizations after the cell are made linear using a second $\lambda/4$ waveplate, and the beams are separated using another PBS. The probe beam alone is detected using a photodiode; therefore, the photodiode signal is proportional to probe transmission. Since the SAS signal used for locking corresponds to absorption by zero-velocity atoms, detecting the non-scanning probe beam allows us to have a flat Doppler-free background for the CBS signal.

\section{CBS in a pure cell}
\subsection{Experimental results}

\begin{figure}[!h]
 	\centering
 	\includegraphics[width=.5\textwidth]{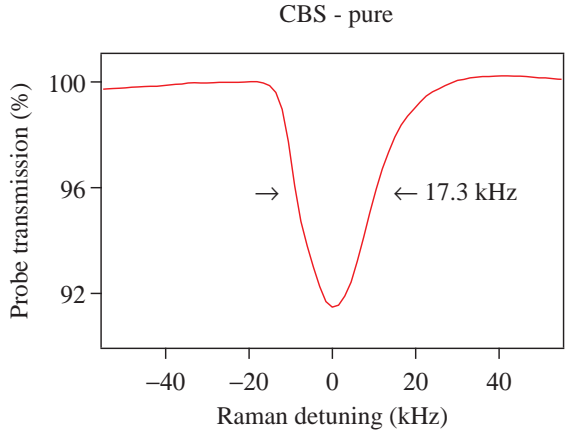}
 	\caption{(Color online) CBS resonance obtained in a pure cell.}
 	\label{cbspure}
 \end{figure}

An experimental spectrum for CBS in the $F_g = 2 \rightarrow F_e = 3 $ transition obtained in a pure cell is shown in Fig.\ \ref{cbspure}. Probe transmission as a function of detuning of the pump beam shows a dip---the CBS resonance---at line center; the photodiode signal is scaled so that the percentage absorption is about 8\%. This behavior is opposite to the CPT resonance seen in the $F_g = 1 \rightarrow F_e = 1 $ transition in the same isotope \cite{KKB17}. The difference is because the $ 1 \rightarrow 1 $ transition does not satisfy the requirements for a CBS resonance (mentioned earlier).

\subsection{Theoretical analysis}

The experimental spectrum can be explained from a detailed density-matrix analysis of the sublevel structure for this transition. The calculations were carried out using the atomic density matrix (ADM) package written by Simon Rochester \cite{ROCadm}. It solves numerically the following time evolution equation for the density-matrix elements involved:
\begin{equation}
\dot{\rho} = - \dfrac{i}{\hbar}[H,\rho] - \dfrac{1}{2}\{\Gamma, \rho \} + \text{repopulation terms}
\end{equation}
where $ \Gamma $ is the relaxation matrix---its diagonal terms gives the total decay rate (radiative and non-radiative) of the respective populations, and its off-diagonal terms represent the decoherence between states $ \ket{i} $ and $ \ket{j} $, such that
\begin{equation}
\Gamma_{ij} = \dfrac{\Gamma_{ii} + \Gamma_{jj}}{2}
\end{equation}
The repopulation terms take care of decay of atoms from the excited state to the ground state.

\begin{figure}[!h]
  	\centering
  	\includegraphics[width=.8\textwidth]{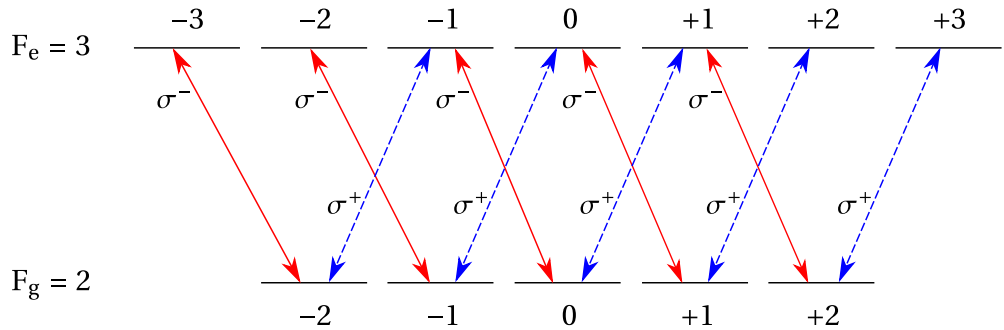}
  	\caption{(Color online) Magnetic sublevels of $F_g=2 \rightarrow F_e=3$ transition in the D$_2$ line of $^{87}$Rb.}
  	\label{cbslevels}
  \end{figure}

The magnetic sublevel structure for the transition is shown in Fig.\ \ref{cbslevels}. The pump beam is $ \sigma^- $ polarized---hence it couples sublevels with the selection rule $ \Delta m = -1 $. The probe beam is $ \sigma^+ $ polarized and couples sublevels with the selection rule $ \Delta m = +1 $.  The probe beam has no detuning for zero-velocity atoms while the pump beam has a detuning for the same atoms, but the actual detuning seen in the atom's frame depends on its velocity.

The following parameters are input to the calculation:
\begin{itemize}
	\item[(i)] The $ F $ values for the ground and excited state of the transition.
	
	\item[(ii)] The proper polarizations for the probe and pump beams.
	
	\item[(iii)] A uniform intensity, equal for both beams.
	
	\item[(iv)] A decay rate among ground sublevels of 10 kHz.
	
	\item[(v)] A decay rate from an excited sublevel to a ground sublevel of 6 MHz.
	
	\item[(vi)] A repopulation term for a particular ground sublevel equal to the 6 MHz decay rate multiplied by the appropriate branching ratio.
\end{itemize}
The probe transition spectrum is Doppler averaged over atomic velocities corresponding to the Maxwell-Boltzmann distribution for Rb atom at room temperature.

\begin{figure}[!h]
   	\centering
    \includegraphics[width=.5\textwidth]{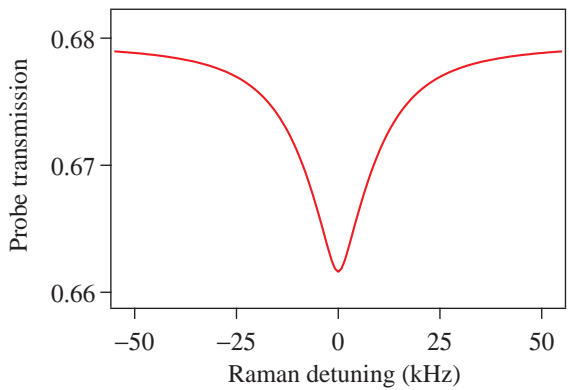}
   	\caption{(Color online) Simulated probe transmission spectrum versus Raman detuning for $F_g=2 \rightarrow F_e=3$ transition with probe and pump beam intensity 21 \textmu W/cm$^2$.  }
   	\label{cbssim}
\end{figure}

The results of the simulation are shown in Fig.\ \ref{cbssim}. The calculated spectrum reproduces the experimental one quite well, in terms of linewidth. The only difference is that the calculation assumes a constant intensity of 21 \textmu W/cm$^2$, which only appears in the wings of the Gaussian distribution for the 30 \textmu W power used in the experiment.

\subsection{Effect of pump power}

In a CBS experiment---like in CPT---the pump beam causes decoherence through the excited state. The scattering rate is intensity dependent, and is given by 
\begin{equation}
R = \dfrac{\Gamma}{2} \dfrac{I/I_s}{1 + I/I_s}
\end{equation}
where $ \Gamma $ is the natural linewidth of the state, $ I $ is the intensity, and $ I_s $ is the saturation intensity (the intensity at which the transition gets power broadened by a factor of $ \sqrt{2} $). Since the intensity is directly proportional to the power through a geometric factor, $ I=g P $, the scattering rate can be rewritten as 
\begin{equation}
R = \dfrac{\Gamma}{2} \dfrac{gP/I_s}{1 + gP/I_s}
\label{scrate}
\end{equation}
This equation shows that the scattering rate will increase initially but asymptote to a saturation value at high powers. Thus the linewidth of the CBS resonance will also show the same behavior.

\begin{figure}[!h]
 	\centering
 	\includegraphics[width=.5\textwidth]{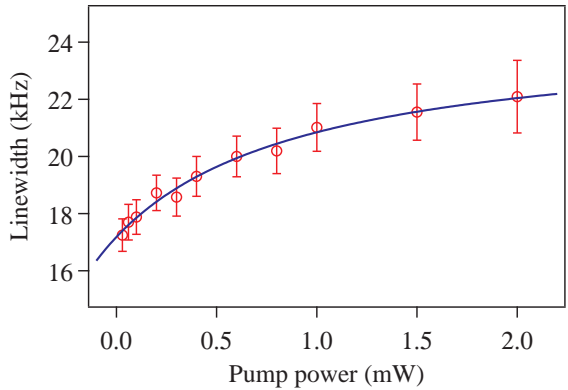}
 	\caption{(Color online) Effect of pump power on the linewidth of the CBS resonance showing increase in linewidth due to increased decoherence through the upper level. The solid line is a fit to the scattering-rate expression in Eq.\ \eqref{scrate}}
 	\label{cbspowervar}
 \end{figure}

The results are shown in Fig.\ \ref{cbspowervar}. The solid line is a fit to Eq.\ \eqref{scrate}, with an offset to account for linewidth from experimental noise. The fit describes the experimental results quite well.

\section{CBS in a buffer cell}
\begin{figure}[!h]
 	\centering
    \includegraphics[width=.5\textwidth]{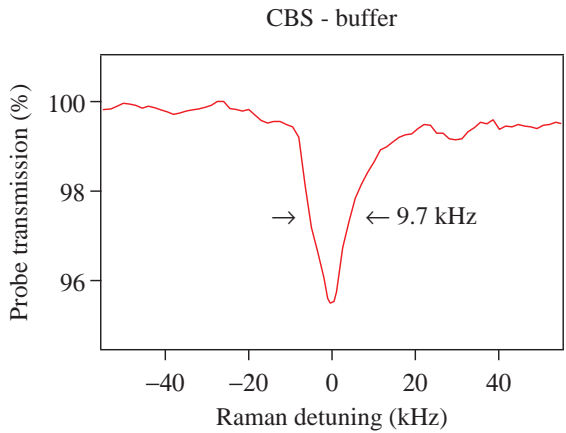} 
 	\caption{(Color online) CBS resonance obtained in a buffer gas filled cell.}
 	\label{cbsbuffer}
 \end{figure}

Before concluding, we turn to experimental results in a buffer cell---one filled with 20 torr of Ne as buffer gas. The role of the buffer gas is to increase the coherence time among magnetic sublevels of the ground state. This will result in a smaller linewidth for the CBS resonance. The results shown in Fig.\ \ref{cbsbuffer} bear out this expectation---the linewidth reduces to 9 kHz in such a cell. In this case, the absorption is a factor of 2 lower than that in a pure cell, and the photodiode signal is scaled to reflect this.

\section{Conclusions}
In summary, we have studied an enhanced absorption or CBS resonance in a closed transition in room-temperature vapor of $^{87}$Rb atoms. The observation requires the proper superposition state to be formed, which is achieved by using magnetic sublevels of the ground state and a phase coherence between the probe and pump beams  by deriving them from the same laser. The observed linewidth is limited by decoherence among the magnetic sublevels. This explanation is borne out by a density-matrix analysis of the sublevels involved in the transition. The calculation takes into account Doppler averaging in room temperature $^{87}$Rb vapor. We study the effect of pump power on the CBS linewidth, and find that the behavior follows a power-dependent scattering rate from the excited state. We also study the same CBS resonance in a buffer-gas filled cell, and find that it reduces the linewidth because it increases the coherence time among the magnetic sublevels.

\section*{Acknowledgments}
This work was supported by the Department of Science and Technology, India. S K acknowledges financial support from INSPIRE Fellowship, Department of Science and Technology, India. The authors thank S Raghuveer for help with the manuscript preparation; and Harish Ravi and Mangesh Bhattarai for helpful discussions.


%

\end{document}